\documentclass[12pt]{iopart}

\usepackage{iopams}
\usepackage{graphicx}
\usepackage{cite}
\newtheorem{lemma}{Lemma}
\begin{document}

%\begin{frontmatter}
% Title, authors and addresses

% use the thanksref command within \title, \author or \address for footnotes;
% use the corauthref command within \author for corresponding author footnotes;
% use the ead command for the email address,
% and the form \ead[url] for the home page:
% \title{Title\thanksref{label1}}
% \thanks[label1]{}
% \author{Name\corauthref{cor1}\thanksref{label2}}
% \ead{email address}
% \ead[url]{home page}
% \thanks[label2]{}
% \corauth[cor1]{}
% \address{Address\thanksref{label3}}
% \thanks[label3]{}

\title{New Parametric Approach for the General Lorenz System}

% use optional labels to link authors explicitly to addresses:
% \author[label1,label2]{}
% \address[label1]{}
% \address[label2]{}

\author{Lazhar Bougoffa$^{1}$ and Smail Bougouffa$^{2}$}
%\corauth[]{Corresponding author}
\address{$^{1}$Al Imam Mohammad Ibn Saud Islamic University
(IMSIU), Faculty of  Science, Department of Mathematics, P.O. Box
90950, Riyadh 11623, Saudi Arabia\\
$^{2}$Department of Physics, Faculty of
Science, Taibah University, P.O. Box 30002, Madina, Saudi Arabia}
\ead{$^{1}$bougoffa@hotmail.com, $^{2}$sbougouffa@hotmail.com and
sbougouffa@taibahu.edu.sa}
\begin{abstract}
We are concerned with the study of the system of coupled equations
of motion for a system of two-level atoms interacting with a
single-mode field in the laser cavity on resonance. The passage
from the equation of motion to the Lorenz equations is
established. Some theoretical aspects of the slowly time -varying
phenomena in cavities are discussed and the linear stability
analysis is presented. A new approach to solve this nonlinear
coupled differential equations is explored and its generalization
to the case of variable coefficients is performed.
\end{abstract}

 \vspace{2pc}


\maketitle

\section{Introduction}\label{1}
In many problems in quantum optics, dynamical chaos plays an
important role. These include, for example, nonlinear systems and
deterministic chaos in lasers. In the last case, it has been shown
that, even when the laser is treated semiclassically by
deterministic equations, the equations are capable of exhibiting
quasi-periodic and even chaotic behavior, but this has nothing to
do with quantum fluctuations \cite{U63, BC64, RN68}. In general,
the study of the time-varying phenomena in cavities shows how
various cavity instabilities can be lead to dynamical chaos, which
occurs in many coupled nonlinear problems ~\cite{MS99,MW95,FW04, H85}.
One of the simplest but popular models in dynamical chaos is the
Lorenz model, which illustrates one of the powerful aspects of the
study of chaos: generic features that are learned from one field -
here hydrodynamics- can readily be adapted to another - laser
physics. Haken ~\cite{H75} showed the existence of  an isomorphism
between the Bloch equations and the Lorenz equations. Of course,
the Lorenz system is of considerable interest in its own right.
This has been true from the beginning when it used in the chaos
problems ~\cite{L63} nearly forty years ago, to a recent emergence
of a computer assisted proof that such comportment is present for
this system ~\cite{MM95}. For more on the Lorenz system, see
~\cite{GH93, S82} and references therein. We are interested,
however, in the evolution of the variable associated with the
nonlinear systems. This requires us to obtain the equations of
motion for the nonlinear systems of interest and solve them. In
order to treat these nonlinear equations and their stability of
their solutions, we briefly present the correspondence between the
Bloch equations for a system of identical two-level atoms
interacting with a single-mode field in the laser cavity on
resonance and the Lorenz equations in hydrodynamic. Then we
discuss some of the characteristics of solutions of these
nonlinear coupled differential equations equations which can not
always be exactly separated. It may, however, be worthwhile if the
physical models can be constructed in such manner that the coupled
system can either be solved analytically or transformed into
another system in which the equations are decoupled and solved
separately. In this work we attempt to present a new analytic
approach to solve this system and then discus the possibility to
extend this approach  for the case of variable coefficients.
Within this scope this paper begins in section 2 with setting up
the Bloch equations for a system of two-level atoms interacting
with a single-mode field in the laser cavity on resonance. In
section 3 we establish the relationship to the Lorenz equations.
Section 4 presents some aspects and features of the Lorenz
equations. Section 5 shows the linearization approach where the
system can be treated using the separation approach discussed in
\cite{SS08}. In section 6, we explore a new approach to solve
analytically the Lorenz system for a particular class of initial
conditions. In Section 7, we  extend our approach to include the
solutions of the Lorenz system with variable coefficients, and we
conclude in last section with some remarks and possible extension
of the domain of investigation.

\section{The single Mode Laser Equations }\label{2}
In the previous papers \cite{SS08, SS07} we have explored various
aspects of the optical Bloch equations (OBE) which are widely used
for describing dynamics in a system consisting of molecules,
electromagnetic waves, thermal bath, and the near-resonant
interaction of spin or atomic transition with radiation field. In
general, the optical Bloch equations are treated in steady-state,
to provide the mean radiative forces for a two- and three-level
atom. Contrary to the common steady state approximation, the
dynamics and transient optical effects on atomic motion can be
considered by solving the complete system in some circumstances.
In the framework of the semiclassical approximation and with the
rotating wave approximation ~\cite{MS99, MW95}, for the pure
radiative decay which is defined by an upper-to-lower-level
transition with the dipole constant $ 1/T_{1}$, the motion
equations of the components of the Bloch vector $(U, V, W)$ form a
system of coupled differential equations. For central tuning, it
can be assumed that $U=0$ and the two remaining equations are
coupled to slowly-varying amplitude equation. Then the system can
be read as
\begin{eqnarray}\label{1, 2, 3}
% \nonumber to remove numbering (before each equation)
   \dot{V} &=& - \frac{1}{T_{2}}V - \Omega(\textbf{r})W, \nonumber \\
  \dot{W} &=&  \Omega({\textbf{r}})V  - \frac{1}{T_{1}}(W +1), \\
  \dot{\Omega}&=& -\frac{1}{T_{c}}\Omega+\omega gV,\nonumber
\end{eqnarray}
where for simplicity, we assume that the field frequency $\omega$
is resonant with the atomic transition $\omega_0$, i.e., the
detuning $\delta=0$ and we lump all rapid spatial dependence of
the field into the Rabi frequency $\Omega(\textbf{r})$ that is now
a function of position in general. The "dot" denotes the time
derivative $d/dt$, $\frac{1}{T_{c}}$ is the cavity damping rate
and $g$ is the
coupling rate between the cavity field and the atom.
In keeping with the literature on NMR we use $T_2$ for the induced dipole decay
and $T_1$ for the probability difference decay time. \\
The usual laser threshold, in which the gain rate equals the loss
rate, matches to the condition
\begin{equation}\label{4}
    \omega g=\frac{1}{T_{c}T_{2}}.
\end{equation}
These coupled equations are known in literature \cite{FW04} as the
Lorenz-Maxwell equations and form a nonlinear system which can not
always be exactly separated. Their nonlinear character can lead to
chaotic instabilities (classical chaos) in atomic and field
dynamics. On the other hand, this system is isomorphic to the
Lorenz equations as it was demonstrated by Haken \cite{H85,H75},
so that a laser governed by these equations is capable of
exhibiting the same chaotic behavior. In the following section we
shall present the passage form these laser equations to the Lorenz
equations, which demonstrate that much more complexity is possible
once we have three coupled variables.

\section{Passage to the Lorenz equations}\label{3}
The original purpose of the Lorenz equations was to introduce a
simple model of the convection instability, which occurs when a
fluid is heated from below and kept at a constant temperature from
above ~\cite{MS99}. For small temperature gradients, the fluid
remains quiescent, but it starts a macroscopic motion as the
gradient approaches a critical value. The heated parts of the
fluid expand, move up by buoyancy, cool and fall back to the
bottom: a regular spatial pattern appears out of a completely
homogeneous state.\\
The motion of the fluid was described by the nonlinear, partial
differential Navier-Stockes equations of hydrodynamics, which are
first undertaken in ~\cite{H52, F73}. In later, by introducing a
special Fourier decomposition of the velocity and temperature
fields in the fluid, Lorenz derived a set of truncated
dimensionless equations coupling just one component $x$ of the
velocity field to two components $y$ and $z$ of the temperature
field ~\cite{H85, FJKT01}. The Lorenz equations have the following
form:
\begin{eqnarray}\label{5,6,7}
% \nonumber to remove numbering (before each equation)
   \dot{x} &=& -ax + ay, \nonumber \\
  \dot{y} &=& -y -xz + rx, \\
 \dot{z} &=& -bz +xy, \nonumber
\end{eqnarray}
where $b>0$, $a >0,$ and $r>1$. The constants $a, r, b$
are related to the physical parameters. For example, in fluid
dynamics $a$ is the Prandtl number, $r=R/R_{c}$, where $R$
is the Rayleigh number and $R_{c}$ the critical Rayleigh number
~\cite{MS99}. $b=4\pi^2/(\pi^2+k_{1}^2)$, where $k_{1}$ is a
dimensionless wave number. These equations are ordinary
differential equations and hold within two nonlinearities in the
form $xz$ and $xy$. To the great surprise of many mathematicians
and physicists these equations can have solutions which are quite
irregular. These solutions were found numerically.\\
In order, to establish the correspondence between the Lorenz
equations (\ref{5,6,7}) and the single-mode laser equations
(\ref{1, 2, 3}), we find the steady-state solutions of Eqs.
(\ref{1, 2, 3}) which can be obtained by putting the time
derivatives equal to zero. We then determine the following
solutions in which the field is non-zero,
\begin{eqnarray}\label{8,9,10}
% \nonumber to remove numbering (before each equation)
  W_{ss} &=&  -\frac{1}{\omega gT_{c}T_{2} }, \nonumber\\
  V_{ss} &=& \frac{1}{\omega gT_{c}}\sqrt{\frac{1}{T_{1}T_{2}}(\omega g T_{c}T_{2}-1)}, \\
  \Omega_{ss} &=& \sqrt{\frac{1}{T_{1}T_{2}}(\omega g
  T_{c}T_{2}-1)},\nonumber
\end{eqnarray}
provide that $\omega g T_{c}T_{2}\geq 1$. It is convenient now to
introduce normalized variables in terms of these steady-state
expressions, and make the following substitution
\begin{equation}\label{11}
    x= \kappa \Omega /\Omega_{ss},\quad\quad y=\kappa V/V_{ss}, \quad\quad z=-(W+1)/W_{ss},
\end{equation}
where
\begin{equation}\label{12}
    \kappa= \sqrt{\frac{T_{2}}{T_{1}}(\omega gT_{c}T_{2}-1)}.
\end{equation}
We attain exactly the equations (\ref{5,6,7}), except that now
\begin{equation}\label{13}
    a= \frac{T_{2}}{T_{c}}, \quad\quad b=\frac{T_{2}}{T_{1}}, \quad\quad r=
    \omega gT_{c}T_{2}
\end{equation}
and time is now in units of $T_{2}$. We note here that these
equations can give a detailed account of turbulence motion and
chaos when $a > b+1$. This condition correspond to the
following condition on the damping constants
\begin{equation}\label{14}
    \frac{1}{T_{c}}> \frac{1}{T_{2}}+\frac{1}{T_{1}},
\end{equation}
which indicates that the atomic decay rates can not exceed the
cavity rate, which is in contradiction with the usual condition
that was assumed in semiclassical theory of laser in order to make
an adiabatic elimination of the atomic variables. It is now clear
that chaos can not show up in semiclassical treatment of laser
problem; it requires a bad cavity, in which the spectral width of
the cavity field exceeds the atomic homogeneous linewidth
\cite{MW95}. The parameter $\omega gT_{2}$ represents the  gain
rate of the laser, which equals the loss rate $1/T_{c}$ at the
laser threshold. Then the quantity $\omega gT_{2}T_{c}$ measures
how many times the gain exceeds the loss. In the following we
shall pass to present some features of Lorenz equations.

\section{Some features of Lorenz equations}\label{4}
We are not concerned directly here with the subject of fluid flow,
but we shall present some important features of the Lorenz
equations which are isomorphic to the single mode laser equations.
In addition to they are nonlinear coupled equations, sensitive to
the initial conditions and all their trajectories are eventually
confined to paths of zero volume, these equations have some
important theoretical aspects, which can briefly summarized in the
following points:\\
A steady state of a system  is a point in phase space from which
the system will not change in time, once that state has reached.
In other words, it is a point, $(x, y,z)$, such that the solution
does not change, or where
\begin{equation}\label{15}
  \dot{x}=0, \quad \dot{y}=0 \quad and \quad
   \dot{z}=0.
\end{equation}
This point is usually referred to as a stationary point of the
system and is called a fixed point. By setting the time
derivatives equal to zero in the Lorenz equations (13, 14, 15) ,
and solving the resulting system we find three possible steady
states, namely the stationary points
\begin{equation}\label{16}
   C^{0}=(0,0,0),\quad C^{\pm}=(\pm \sqrt{b(r-1)}, \pm \sqrt{b(r-1)}, r-1),
\end{equation}
so that there is only one fixed point when $0 \leq r \leq 1$ but
all three fixed points are present when $r>1$ which corresponds to
the laser threshold \cite{MS99}. As $r\rightarrow 1$, $C^{\pm}$
coalesce with the origin in a pitchfork bifurcation \cite{S82}.\\
The stability of the non-linear system is almost always similar to
that the linearized systems near the fixed points. Then, the
stability of the fixed points of the Lorenz equations can be
established by linear stability analysis for more details see
~\cite{S82,YC04}. The origin (0,0,0) is a stable fixed point for
$r < 1$; that is it attracts nearby solutions to itself and it
becomes unstable for $r > 1$. In this last case the two other
fixed points in turn become unstable when $a > b+1$ and $r >
a (a +b+3)/(a -b -1)$. This is called a Hopf
bifurcation ~\cite{YC04, SSLM88}.\\
The right-hand side of the second inequality can be considered as
a function of $a$ and it has a minimum when
\begin{equation}\label{17}
    a = 1+b+\sqrt{2(b+1)(b+2)},
\end{equation}
The minimum value of the function can be given by
\begin{equation}\label{18}
    \min. = 2\sqrt{2(b+1)(b+2)}+3(b+1)+2,
\end{equation}
which depends only on the parameter $b$.\\
In the laser case, this minimum value represents the minimum value
of the gain-to-loss-ratio, which shows that how large this has to
be far for instability to be encountered, and reads
\begin{equation}\label{19}
    \min. =
    2\left[2\left(\frac{T_{2}}{T_{1}}+1\right)\left(\frac{T_{2}}{T_{1}}+2\right)\right]^{1/2}
    + 3\left(\frac{T_{2}}{T_{1}}+1\right) +2,
\end{equation}
The minimum gain-to-loss ratio is refereed to as the second laser
threshold, depends on $T_{2}/T_{1}$, but always exceeds 9.\\
The difficulty part of doing any theoretical analysis of the
Lorenz equations is that they are non-linear. So, the system of
non-linear equations can be approximated by a linear one, using a
Taylor series around a fixed point. The basic idea is to replace
the right hand side functions in the equations (\ref{5,6,7}) with
linear approximation about the fixed point, and then solve the
resulting system of linear ordinary differential equations . The
obtained linear system can be treated using our technique that
discussed in references ~\cite{SS07, SS08}.

\section{Linear stability treatment}\label{5}
In order to examine whether the steady-state solution for the
fixed point given by (\ref{15}) represents a stable state, we linearize
the Lorenz equations in the neighborhood of the fixed point by
using Taylor series of functions of multi-variables: $f(x, y, z)$.
If we linearize a function  $f(x, y, z)$ about $(x_{s},
y_{s},z_{s})$ we obtain the approximation
\begin{eqnarray}\label{20}
   f(x, y, z)& \approx & f(x_{s},
y_{s},z_{s})+f_{x}(x_{s}, y_{s},z_{s})(x-x{s}) + {} \nonumber
\\ & & {} + f_{y}(x_{s}, y_{s},z_{s})(y-y_{s})+f_{z}(x_{s},
y_{s},z_{s})(z-z_{s}).
\end{eqnarray}
Applying this relation to the right hand side function for each of
the ordinary differential equations in (\ref{5,6,7}),then the
linearization of the original equations about the origin $C^{0}$
yields
\begin{eqnarray}\label{21,22,23}
% \nonumber to remove numbering (before each equation)
  \dot{x}&=& -a x + a y ,\nonumber\\
 \dot{y} &=& -y +rx ,\\
 \dot{z} &=& -bz ,\nonumber
\end{eqnarray}
Hence, the z-motion decouples, leaving the two first equations
linearly coupled. The solution to these equations is given by
\begin{equation}\label{24}
% \nonumber to remove numbering (before each equation)
  h(t)=C_{h1}e^{\lambda_{1}t}+C_{h2}e^{\lambda_{2}t}+C_{h3}e^{\lambda_{3}t},
  \quad
  h=x, y, z.
\end{equation}
The $C_{hi}$'s are constants that are determined by the initial
conditions of the problem, and $\lambda_{i}$ are the eigenvalues
of the matrix of the coefficients of x,y, and z in the right hand
side of equations (\ref{21,22,23}), which are given by
\begin{equation}\label{25}
\lambda_{1}=-b,\quad \textrm{and} \quad
\lambda_{2},\lambda_{3}=\frac{1}{2}\left(-a-1\pm
\sqrt{(a-1)^2+4a r}\right).
\end{equation}
The stability of the non-linear system is almost always similar to
that the linearized systems near the fixed points ~\cite{SB08}. Then, the
stability of the fixed points of the Lorenz equations can be
established by linear stability analysis for more details see
~\cite{S82,  YC04}. For $r < 1$ all directions are incoming. The
origin (0,0,0) is a sink and stable node. If $r > 1$, it attracts
nearby solutions to itself and it becomes a 2D saddle point and
unstable: two incoming and one outcoming directions. In this last
case the two other fixed points in turn become unstable when
$a > b+1$ and $r > r_{t}= a (a +b+3)/(a -b
-1)$. This is called a Hopf bifurcation ~\cite{YC04}. The
bifurcations occur at:
\begin{itemize}
\item[*] $r=1$, when the origin changes from stable to unstable, and
two other fixed points appear.
\item[*] $ r=r_{t}$, where the
remaining two fixed points change from being stable to unstable.
\end{itemize}
When $r>1$, the same linearization process can be applied at the
remaining two stationary points $C^{\pm}$,which give an other
system of equations:
\begin{eqnarray}\label{26,27,28}
% \nonumber to remove numbering (before each equation)
  \dot{x}&=& -a x + a y ,\nonumber\\
 \dot{y} &=& -y +x\pm\sqrt{b(r-1)}\overline{z} ,\\
 \dot{\overline{z}} &=& \pm\sqrt{b(r-1)}(x+y) -b\overline{z} -2b(r-1),\nonumber
\end{eqnarray}
where $\overline{z}=z-r+1$. These equations are similar to those
of Bloch studied in \cite{SS08}. The same treatment can be use
here to obtain the solutions of this linear system. The
eigenvalues of this system satisfy another characteristic
equation:
\begin{equation}\label{29}
    \lambda^{3}+(a+b+1)\lambda^{2}+(r+a)b\lambda+2a
    b(r-1)=0.
\end{equation}
In fact, the eigenvalues give us all the information we need to
know about how the linearized solution behaves in time. It is
possible that two of the eigenvalues can be complex numbers that
if $\lambda_{2}=\lambda_{3}=\alpha\pm i\beta$ then the homogeneous
solutions can be rearranged so that they are of the form
\begin{equation}\label{30}
% \nonumber to remove numbering (before each equation)
  h(t)=C_{h1}e^{\lambda_{1}t}+C_{h2}e^{\alpha t}\cos(\beta t)+C_{h3}e^{\alpha t}\sin(\beta
  t),
  \quad
  h=x,y,\overline{z}.
\end{equation}
In terms of the asymptotic stability of the problem, we need to
look at the asymptotic behavior of the solutions (\ref{30}), as
$t\rightarrow \infty$, from which several conclusions can be
drawn:
\begin{enumerate}
 \item If the eigenvalues are real and at least one is
positive, then the solutions will blow up as $t\rightarrow
\infty$. In this case
the linearized solutions are unstable.\\
\item If the eigenvalues are real and negative, then the solutions
will go to zero as $t\rightarrow \infty$. In this case the
linearized solutions are stable.\\
\item If there is a complex conjugate pair of eigenvalues, $\alpha\pm i\beta$,
then the solutions exhibit oscillatory behavior (with the
emergence of the terms $sin(\beta t)$ and $cos(\beta t)$). If the real part,
$\alpha$, of all eigenvalues is negative, the oscillations will decay
in terms of time and the solutions are stable; if $\alpha>0$, then the
oscillations will grow, and the solutions are unstable.\\
\item If the complex eigenvalues are pure imaginary, then the
oscillations will neither increase nor decay in terms of time. The
linearized solutions are periodic, and marginally stable.\\
\end{enumerate}
It is a clear that the linear results apply only near the
stationary points, and do not apply to all of the phase space. One
difficulty that has not mentioned is that for values of $r>r_{t}$,
the systems have oscillatory solutions, which are unstable. Linear
theory does not reveal what happens when these oscillations become
large. In order to study more closely the long-time behavior of
the solution, we must recourse to another approach, which will be
presented in the following parts to avoid the numerical
treatments \cite{LSS07}. From the point of view experimental, Klische and Weiss
\cite{KW85} achieved in observing a sequence of pulsations of the
light output, including successive period doubling and eventually
chaos. Furthermore, Harrison et al. \cite{HSB85} succeeded in
observing of instabilities leading to chaos in the emission from
single mode homogeneous broadened
single mode and multi-mode midinfrared Raman laser.

\section{Construction of exact parametric solutions of the Lorenz system}\label{6}
In this section, we will present a direct approach to solve  the
Lorenz model with constant coefficients.\\ Multiplying both sides
of the first equation of system (\ref{5,6,7}) by $x$ we can get
\begin{equation}\label{31}
xy=\frac{1}{a}\left[x\frac{dx}{dt}+ax^{2}\right].
\end{equation}
Substituting Eq. (\ref{31}) into the third equation of system
(\ref{5,6,7}), we can obtain
\begin{equation}\label{32}
\frac{d}{dt}\left(z-\frac{1}{2a}x^{2}\right)=
-b\left(z-\frac{1}{b}x^{2}\right).
\end{equation}
Assuming that   $b=2a,$ thus  Eq. (\ref{32}) gives
\begin{equation}\label{33}
z-\frac{1}{b}x^{2}= ce^{-bt},
\end{equation}
where $c$ is a free parameter (constant of integration) that can
be chosen equal zero for some class of
initial conditions $z(0)-\frac{1}{b}x^2(0)=0$.\\
Now, substituting Eq. (\ref{32}) into the second differential
equation of system (\ref{5,6,7}), we obtain
\begin{equation}\label{34}
\frac{dy}{dt}=rx-y-\frac{1}{b}x^{3}.
\end{equation}
Differentiating both sides of the first equation of system
(\ref{5,6,7}) with respect to $t,$ we get
\begin{equation}\label{35}
\frac{d^{2}x}{dt^{2}}=-a\frac{dx}{dt}+a\frac{dy}{dt}.
\end{equation}
Substituting Eq. (\ref{34}) into Eq. (\ref{35}), taking into
account that $b=2a,$ and in view  of
\begin{equation}\label{36}
ay=ax+\frac{dx}{dt},
\end{equation}
we have
\begin{equation}\label{37}
\frac{d^{2}x}{dt^{2}}+(a+1)\frac{dx}{dt}+a(1-r)x=-\frac{1}{2}x^{3}.
\end{equation}
Let $v=\frac{dx}{dt}.$ Since
\begin{equation}\label{38}
\frac{d^{2}x}{dt^{2}}=\frac{dv}{dt}=\frac{dv}{dx}\frac{dx}{dt}=v\frac{dv}{dx}
\end{equation}
Thus
\begin{equation}\label{39}
v\frac{dv}{dx}=-(a+1)v+a(r-1)x-\frac{1}{2}x^{3}
\end{equation}
This is the Abel differential equation of the second kind. \\
The substitution $\xi=-(a+1)x$ guides the Abel equation to the
simplified form
\begin{equation}\label{40}
v\frac{dv}{d\xi}-v=\phi(\xi),
\end{equation}
where the function $\phi(\xi)$ is defined parametrically by the
relations
\begin{equation}\label{41}
\phi(\xi)=
\frac{a(r-1)}{(a+1)^{2}}\xi-\frac{1}{2(a+1)^{4}}\xi^{3}.
\end{equation}
Once $v$ is found then we can get $x$ through
$\frac{dx}{dt}=v(x).$\\
Much more about the Abel equations and a large number of their
solutions for various $\phi(\xi),$ can be found in standard
textbooks, e.g. ( Polyamin and Zaitsev \cite{13}). Particularly,
the solution of the following equation
\begin{equation}\label{42}
v\frac{dv}{d\xi}-v=-\frac{2(m+1)}{(m+3)^{2}}\xi+A\xi^{m},
\end{equation}
is given in the parametric form as follows:
\begin{equation}\label{43}
\xi=\frac{m+3}{m-1}\alpha\tau E_{m}^{\frac{2}{m-1}},
\end{equation}
\begin{equation}\label{43}
v=\alpha
E_{m}^{\frac{2}{m-1}}\left(R_{m}E_{m}+\frac{2}{m-1}\tau\right),
\end{equation}
where
\begin{equation}\label{44}
A=\pm\frac{m+1}{2}\left(\frac{m-1}{m+3}\right)^{m+1}\alpha^{1-m},
\end{equation}
\begin{equation}\label{45}
 E_{m}=\int\left(1\pm\tau^{m+1}\right)^{\frac{-1}{2}}d\tau-C_{1}
\end{equation}
and
\begin{equation}\label{46}
R_{m}=\sqrt{1\pm\tau^{m+1}}.
\end{equation}
Here $\tau$ is a parameter and $C_{1}$ is an arbitrary constant.\\
Combining Eqs. (\ref{40}, \ref{41}) with Eqs. (\ref{42}-\ref{46}),
we obtain
\begin{equation}\label{47}
\frac{a(r-1)}{(a+1)^{2}}=-\frac{2(m+1)}{(m+3)^{2}}
\end{equation}
and
\begin{equation}\label{48}
\frac{1}{2(a+1)^{4}}=\frac{m+1}{2}\left(\frac{m-1}{m+3}\right)^{m+1}\alpha^{1-m},
\end{equation}
where $m=3.$ After some algebra, we obtain
\begin{equation}\label{49}
r=1-\frac{\alpha}{a} \  \mbox{and} \ \alpha=\pm
2\left(\frac{a+1}{3}\right)^{2}.
\end{equation}
Therefore our result can be reformulated as
\begin{lemma}\label{L1}
If $b=2a$  and $r=1-\frac{ \alpha}{a}, $ where $\alpha=\pm
2\left(\frac{a+1}{3}\right)^{2}.$ Then the solution
$(x(t),y(t),z(t))$ of system (\ref{5,6,7}) for a class of initial
conditions
$z(0)-\frac{1}{b}x^2(0)=0$ is given as follows:\\
\begin{equation}
\frac{dx}{dt}=v(x),
\end{equation}
where $v$ is given in parametric form by
\begin{equation}
\xi=3\alpha\tau E_{3}, \ \xi=-(a+1)x,
\end{equation}
\begin{equation}
v(\xi)=\alpha E_{3}\left(R_{3}E_{3}+\tau\right),
\end{equation}
where
\begin{equation}
 E_{3}=\int\left(1\pm\tau^{4}\right)^{\frac{-1}{2}}d\tau-C_{1}
\end{equation}
and
\begin{equation}
R_{3}=\sqrt{1\pm\tau^{4}}.
\end{equation}
Once $v$ is found then we can get $x(t)$ through
$\frac{dx}{dt}=v(x),$
\begin{equation}
z(t)=\frac{1}{b}x^{2}(t)
\end{equation}
and $y(t)$ can be obtained from
\begin{equation}
\frac{dy}{dt}=rx-y-xz,
\end{equation}
that is
\begin{equation}
y(t)=e^{-t}\int e^{t}\chi(t)dt+C_{2}e^{-t}, \ \chi(t)=
rx(t)-x(t)z(t).
\end{equation}
\end{lemma}

\section{The general Lorenz system with variable
coefficients}\label{8}
One of the main problem of mathematical
physics appears when $a(t), \ b(t)$ and $r(t)$  are analytic
functions and are added to the original system. The new problem,
incorporating the above functions, is
\begin{eqnarray}\label{50}
\left\{
\begin{array}{rcl}
\frac{dx}{dt}&=&-a(t)x+a(t)y,\\\\
\frac{dy}{dt}&=&r(t)x-y-xz,\\\\
\frac{dz}{dt}&=& -b(t)z+xy.
\end{array}
\right.
\end{eqnarray}
A question which arises naturally is under what conditions on the
functions $a(t), \ b(t)$ and $r(t)$ does the given system have an
explicit solution?\\ In this section, we will propose a similar
approach to solve the
general Lorenz model.\\
Proceeding as before, after some algebra we can get
\begin{equation}\label{51}
\frac{d}{dt}\left(z-\frac{1}{2a(t)}x^{2}\right)=
-b(t)\left(z-\frac{1}{b(t)}\left[
1-\left(\frac{1}{2a(t)}\right)'\right]x^{2}\right).
\end{equation}
where the prime means a first derivative respect to $t$. Assuming
that
\begin{equation}
\frac{1}{b(t)}\left[1-\left(\frac{1}{2a(t)}\right)'\right]=\frac{1}{2a(t)}.
\end{equation}
This means that
\begin{equation}\label{52}
b(t)-2a(t)-\frac{a'(t)}{a(t)}=0.
\end{equation}
Thus Eq. (\ref{51}) gives
\begin{equation}\label{53}
z-\frac{1}{2a(t)}x^{2}= c_{2}e^{-\int b(t)dt},
\end{equation}
where $c_{2}$ is a free parameter (constant of integration). \\
Now, substituting Eq. (\ref{53}) into the second differential
equation of system (\ref{50}), we obtain
\begin{equation}\label{54}
\frac{dy}{dt}=r_{1}(t)x-y-\frac{1}{2a(t)}x^{3},
\end{equation}
where $r_{1}(t)=r(t)-c_{2}e^{-\int b(t)dt}.$\\
Differentiating both sides of the first equation of system
(\ref{50}) with respect to $t$, taking in consideration
Eq.(\ref{54}) and in view of $y=x+\frac{1}{a(t)}\frac{dx}{dt}$, we
obtain
\begin{equation}\label{55}
\frac{d^{2}x}{dt^{2}}+p(t)\frac{dx}{dt}+q(t)x=-\frac{1}{2}x^{3},
\end{equation}
where $p(t)=a(t)+1-\frac{a'(t)}{a(t)}$ and
$q(t)=a(t)\left(1-r_{1}(t)\right).$\\
If we assume that
\begin{equation}\label{56}
a(t)+1-\frac{a'(t)}{a(t)}=0
\end{equation}
and
\begin{equation}\label{57}
a(t)\left(1-r_{1}(t)\right)=c_{3},
\end{equation}
where $c_{3}$ is a constant. \\ Combining Eq. (\ref{52}) with Eq.
(\ref{56}), we get
\begin{equation}\label{58}
b(t)=3a(t)+1.
\end{equation}
On the other hand, Eq. (\ref{56}) read as
\begin{equation}
a'(t)=a(t)+a^{2}(t),
\end{equation}
which is a Bernoulli's equation and its solution reads as
\begin{equation}
a(t)=\frac{c_{4}e^{t}}{1-c_{4}e^{t}}.
\end{equation}
Then Eq. (\ref{55}) becomes
\begin{equation}\label{59}
\frac{d^{2}x}{dt^{2}}+c_{3}x+\frac{1}{2}x^{3}=0.
\end{equation}
Multiplying both sides of Eq. (\ref{59}) by $\frac{dx}{dt}$ and
after some simple algebra we get
\begin{equation}
\frac{d}{dt}\left[ \left(\frac{dx}{dt}\right)^{2}+
c_{3}x^{2}+\frac{x^{4}}{4}\right]=0.
\end{equation}
Thus
\begin{equation}
 \left(\frac{dx}{dt}\right)^{2}+
c_{3}x^{2}+\frac{x^{4}}{4}=c_{5}.
\end{equation}
which can be solved for $\frac{dx}{dt}$, we obtain the separable
differential equation
\begin{equation}
\frac{dx}{dt}=\sqrt{c_{5}-c_{3}x^{2}-\frac{x^{4}}{4}},
\end{equation}
thus,
\begin{equation}\label{60}
\int\frac{dx}{\sqrt{4c_{5}-4c_{3}x^{2}-x^{4}}}=\frac{t}{2}+c_{6}.
\end{equation}
Therefore our result can be reformulated as
\begin{lemma}\label{L2}
If $$b(t)=3a(t)+1$$  and $$r(t)=c_{2}e^{-\int
b(t)dt}+1-\frac{c_{3}}{a(t)}, $$ where
$$a(t)=\frac{c_{4}e^{t}}{1-c_{4}e^{t}}.$$ Then the solution
$(x(t),y(t),z(t))$ of system (\ref{50}) is given as follows:\\
$x(t)$ can be found implicitly by Eq. (\ref{60}), $z(t)$ can be
obtained by Eq. (\ref{53}) and $y(t)$ can be obtained from
\begin{equation}
\frac{dy}{dt}=r(t)x-y-xz,
\end{equation}
that is
\begin{equation}
y(t)=e^{-t}\int e^{t}\chi(t)dt+c_{1}e^{-t}, \ \chi(t)=
r(t)x(t)-x(t)z(t).
\end{equation}
\end{lemma}
\section{Discussion and Conclusion}\label{C}

We have proposed a new method to refine the numerical methods to solve analytically a system of nonlinear coupled differential equations; i.e. the Lorenz equations   by performing a special transformation and then attaining the Abel differential equations that can be solved analytically. The conditions of a total separation form an interesting field of investigation. The separation condition $a=2b$ as it is stated in lemma (1) means that these solutions are not chaotic, since the condition of the existence of the chaotic solutions for the considered system is $a>b+1$. But the analytic form of the non chaotic solutions can be obtained by our proposed approach and the conditions of separation constitute an interesting subject of investigation.
Indeed, the conditions of separation of Lorenz equations as they are stated in lemma (1) are $b=2a$ and $r=1-\alpha/a$, which can be read in the case of the single mode laser equations
\begin{eqnarray}
% \nonumber to remove numbering (before each equation)
  \frac{1}{T_1} &=& \frac{2}{T_c} \\
  w g T_cT_2 &=& \frac{13}{9} +\frac{2}{9}\left(\frac{T_2}{T_{c}}+\frac{T_c}{T_{2}}\right)
\end{eqnarray}
The first equation means that for the probability difference decay time $T_1$ should be half of the cavity damping time $T_c$ that can be realized in the new technology construction of QED-cavity, while the coherence decay time $T_2$ is not conditioned. On the other hand, the second condition shows that the limitation on the coupling rate $g$ between the cavity field and the atom is far for the laser threshold, in which the gain rate equals the loss rate $w g T_cT_2 =1$. However, in the process of dropping the problem from three variables to one variable we also reduce the possibility of encountering certain instabilities that one-dimensional equations cannot reveal. Furthermore, due to some inevitable losses in experiments, the initial conditions that can be practically realized offer an interesting argument in our separation approach. The present procedure shows that the system can be analytically solved for a class of initial conditions \cite{14}, which is given by Eq.(23) that can be read in terms of the parameters of the single-mode laser as
\begin{equation}\label{23}
    -W(0)=\rho_{aa}(0)-\rho_{bb}(0)=1+\kappa\frac{T_2}{T_c}\frac{W_{ss}}{\Omega_{ss}}\Omega(0)
\end{equation}
where $\rho_{aa}, \rho_{bb}$ are the diagonal element of the density matrix that represent the probabilities in the ground and excited states.
Finally, the separation approach of nonlinear coupled equations constitutes an important challenge in mathematics as well as in physics.\\
In general the investigation the conditions, which make the system separable or can be transformed to another system that can be solved analytically, is an interesting point of research. For these reasons, the Lorenz equations attract our intention to explore some analytic solution in order to avoid
and reduce the numerical volume. We have presented a new parametric approach  to solve this nonlinear coupled differential equations and an eventual
extension for the case of variable coefficients.
These results engender a series of questions of mathematical as well as physical consequence. Perfectly one would like to be capable to predict the behavior of paths for any set of initial conditions and parameter values. This is very much an interesting question in general. The Lorenz equations play an important task in the chaotic aspects, especially in laser physics and quantum optics, and have motivated the progress of techniques to explore more and more complicated and higher dimensional models.

\end{document}